\documentclass[a4paper]{article}

\usepackage{INTERSPEECH2019}
\usepackage{url}
\usepackage{color}

\newcommand{\EE}       {\boldsymbol{\Sigma}}
\newcommand{\tmatrix}  {\mathbf{T}}
\newcommand{\mm}       {\boldsymbol{\mu}}

\newcommand{\bphi}{\boldsymbol{\phi}}

\newcommand{\ExpBas}        {$\mathrm{B}$}
\newcommand{\ExpSchOneInit} {$\mathrm{C_{0}}$}
\newcommand{\ExpSchOne}     {$\mathrm{C}$}
\newcommand{\ExpSchTwoInit} {$\mathrm{R_{0}}$}
\newcommand{\ExpSchTwo}     {$\mathrm{R}$}
\newcommand{\ExpFull}       {$\mathrm{D}$}

\title{Factorization of Discriminatively Trained i-vector Extractor for Speaker Recognition}
\name{Ond\v{r}ej Novotn\'y, Old\v{r}ich Plchot, Ond\v{r}ej Glembek and Luk\'a\v{s} Burget\thanks{The work was supported by the Czech Ministry of Interior project No. VI20152020025 ``DRAPAK'', Google Faculty Research Award program, Czech National Science Foundation (GACR) projects No. GJ17-23870Y and ``NEUREM3'' No. 19-26934X, and Czech Ministry of Education, Youth and Sports from the National Programme of Sustainability (NPU II) project ``IT4Innovations excellence in science - LQ1602''.}}

\address{Brno University of Technology, Speech@FIT and IT4I Center of
Excellence, Brno, Czechia}

\email{inovoton@fit.vutbr.cz}

\begin{document}

\maketitle
\begin{abstract}
In this work, we continue in our research on i-vector extractor for speaker verification (SV) and we optimize its architecture for fast and effective discriminative training. 
We were motivated by computational and memory requirements caused by the large number of parameters of the original generative i-vector model. 
Our aim is to preserve the power of the original generative model, and at the same time focus the model towards extraction of speaker-related information.
We show that it is possible to represent a standard generative i-vector extractor by a model with significantly less parameters and obtain similar performance on SV tasks.
We can further refine this compact model by discriminative training and obtain i-vectors that lead to better performance on various SV benchmarks representing different acoustic domains.
\end{abstract}


\section{Introduction}
\label{sec:intro}

In recent years, there have been many attempts to take advantage of neural
networks (NNs) in speaker verification. 
Most of the attempts have replaced or improved one of the components
of an i-vector + PLDA system (feature extraction, calculation of sufficient statistics, i-vector extraction or PLDA) with a neural network. As examples, let us mention: using NN bottleneck features instead of conventional MFCC features \cite{lozano_odyssey_2016}, NN acoustic models replacing Gaussian Mixture Models for extraction of sufficient statistics~\cite{Lei_icassp_2014}, NNs for either complementing PLDA \cite{Novoselov_interspeech_2015,Bhattacharya_SLT16} or replacing it \cite{Ghahabi_icassp_2014}. More ambitiously, NNs that take the frame level features of an utterance as input and directly produce an utterance level representation---usually referred to as an \emph{embedding}---have in the past two years almost replaced the generative i-vector approach in text independent speaker recognition~\cite{Variani_icassp_2014, heighold_icassp_2016, zhang_slt_2016, snyder_slt_2016, Bhattacharaya_interspeech_2017, snyder_interspeech_2017, xvec:Snyder2018}. 

These embeddings are obtained by the means of \emph{pooling mechanism}, for example taking the mean, over the frame-wise outputs of one or more layers in the NN~\cite{Variani_icassp_2014}, or by the use of a recurrent NN~\cite{heighold_icassp_2016}.  An obvious advantage---compared to i-vectors---lies in a much smaller amount of model parameters, which is typically around 10 million in the \emph{x-vector} case~\cite{snyder_interspeech_2017, xvec:Snyder2018} compared to the i-vector with approximately 50 million parameters for both UBM and i-vector extractor. This results in a very fast and memory efficient embedding extraction. A disadvantage of the x-vector framework can be seen in training during which it is essential to massively augment the training data and split them into many rather short (2--5 seconds) examples.

In this work we continue with our research from~\cite{novotny2018discriminatively}, where we kept the large parameter space from the generative i-vector extractor and we focused on discriminative retraining of such a model. We were able to retain the model robustness and even increase the SV performance via optimizing the model for discrimination between speakers---a task closely related to the final speaker verification. However, memory requirements and large computational cost during training have not only limited us in running experiments effectively, but more importantly it was preventing us from continuing with our research goal which is to include this model in a larger DNN scheme that is closer to an end-to-end system.

To solve our problem, we had to drastically decrease the number of trainable model parameters, but, of course, without a major decrease in performance. In the past, people have dealt with the same issue and experimented with factorization of similar or even the same models as ours. In 2003, Subspace Precision and Mean model (SPAM) for acoustic modeling in speech recognition was introduced in \cite{Axelrod2003LargeVC} and later optimized by Daniel Povey in~\cite{PoveySPAM}. SPAM models are Gaussian mixture models with a subspace constraint, where each covariance matrix is represented as a weighted sum of globally shared full-rank matrices. In 2014, Sandro Cumani proposed an i-vector extractor factorization~\cite{sandroFACTOR}, for faster i-vector extraction and smaller memory footprint, where each row of the i-vector extractor matrix is represented as a linear combination of the atoms of a common dictionary with the assumption that it is not necessary to store all rows this matrix to perform i–vector extraction.

In our approach to factorization, we were inspired by~\cite{sandroFACTOR}, but instead of factorizing each row, we perform factorization on the level submatrices of the i-vector extractor that represent individual GMM-UBM components. Also, our motivation is different, as we aim to greatly decrease the memory footprint and therefore substantially speedup the discriminative training. For now, we ignore the possible i-vector extraction speedup. 

To finally obtain a discriminative i-vector extractor, we still use the same strategy as in the x-vector framework~\cite{Variani_icassp_2014, Bhattacharaya_interspeech_2017, snyder_interspeech_2017} and we retrain the NN representation of our factorized generative model to optimize the multi-class cross-entropy over a set of training speakers.  
This is in contrast with our previous research~\cite{dix:glembek}, where we optimized the binary cross-entropy over verification trials formed by pairs of i-vectors. We show that, with such an approach, we can achieve a reasonable performance.  Our results are perhaps not as competitive as those achieved with current state-of-the-art x-vector systems~\cite{xvec:ondran}, nevertheless, we are now closer to our goal which is to further use this model in the fully end-to-end discriminative system~\cite{rohdin:icassp:2018} that can be initialized from a robust generative baseline.

In order to compare both approaches (generative and discriminative) on a speaker verification task, both versions of i-vectors were extracted and used in a standard generative PLDA backend.


\section{Theoretical Background}
\label{sec:SRE}

\begin{figure}[tb]
  \centering
  \scalebox{0.4}{
  \includegraphics[width=1.0\linewidth]{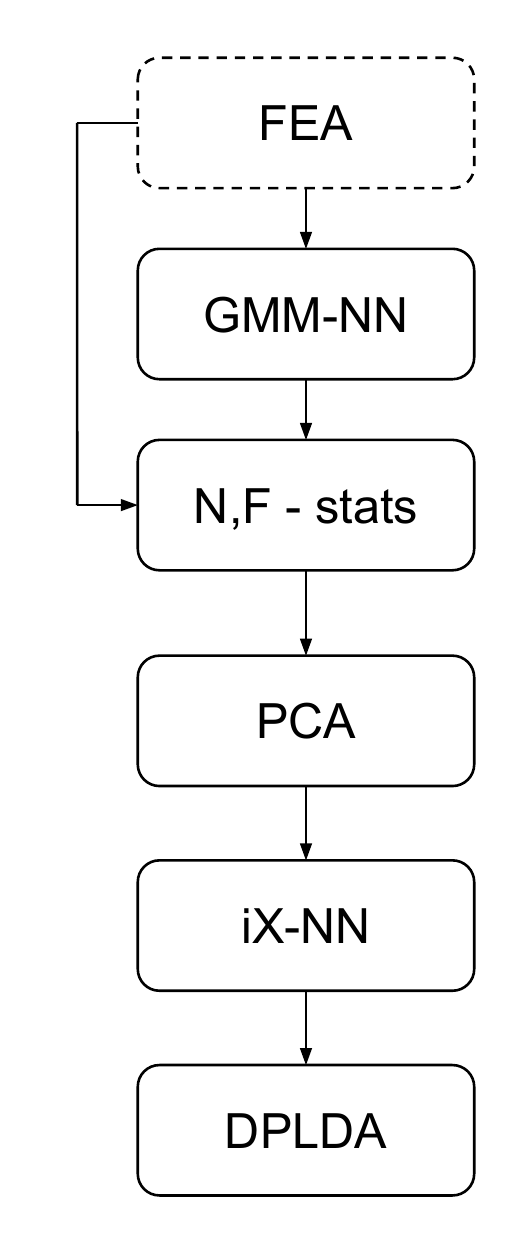}}
  \caption{Scheme of an end-to-end speaker verification system based on a feed forward NN designed to mimic a generic speaker verification system (\cite{rohdin:icassp:2018}).}
  \label{fig:end2end}
\end{figure}
In~\cite{rohdin:icassp:2018}, we had built an end-to-end system (Fig.~\ref{fig:end2end}) that already seemingly fits our goal, but it was exactly the i-vector extractor component that posed the biggest challenge and we had to resort to ad-hoc simplifications, such as PCA-based dimensionality reduction of large dimensional sufficient statistics coming from the GMM-UBM.  
Our approach was to represent a standard generative i-vecror-based SV system as a series of ``elementary'' feed-forward NNs, each representing individual i-vector building block (e.g. GMM-UBM, i-vector extractor,  PLDA classifier).  In the beginning, each NN was trained separately to mimic the equivalent block from the generative training. After this ``initialization", all blocks were connected and jointly retrained.

In this paper, we are focused on the i-vector extractor block and its effective discriminative retraining. We still keep the generative GMM-UBM and PLDA models.


\subsection{i-vector Baseline}

The i-vectors~\cite{DehakN_TASLP:2010} provide a way of reducing
large-dimensional input data to a low-dimensional feature 
vector while retaining most of the relevant information. 
The main principle is that the utterance-dependent Gaussian 
Mixture Model (GMM) supervector of concatenated
mean vectors lies in a low-dimensional subspace---defined by a $CF \times D$ matrix $\tmatrix = [\tmatrix^{(1)}{}', \ldots, \tmatrix^{(C)}{}']'$, commonly referred to as an \emph{i-vector extractor}, with $C$ being number of GMM components, $F$ being feature dimensionality, and $D$ being subspace dimensionality---and whose coordinates are given by the ($D$-dimensional) i-vector $\bphi$. 
The closed-form solution for computing the i-vector can be expressed as a function of
the \emph{zero-} and \emph{first-order GMM statistics}:
$\mathbf{n}_\mathcal{X}=[N_{\mathcal{X}}^{(1)}, \ldots,N_{\mathcal{X}}^{(C)}]'$
and $\mathbf{f}_\mathcal{X}=[\mathbf{f}_{\mathcal{X}}^{(1)'},
\ldots,\mathbf{f}_{\mathcal{X}}^{(C)'}]'$, where
\begin{eqnarray}
\label{Eq:ZeroOrderStats}
N_{\mathcal{X}}^{(c)} & = & \sum_{t} \gamma_{t}^{(c)} \\[-2mm]
\label{Eq:FirstOrderStats}
\mathbf{f}_{\mathcal{X}}^{(c)} & = & \sum_{t} \gamma_{t}^{(c)} \mathbf{o}_{t} ,
\end{eqnarray}
where $\gamma_{t}^{(c)}$ is the posterior (or occupation) probability of frame
$t$ being generated by the mixture component $c$.  
%
The i-vector is then computed as
\begin{equation}
\label{Eq:westimation}
\bphi_{\mathcal{X}} = \mathbf{L}_{\mathcal{X}}^{-1} \bar{\mathbf{T}}' \bar{\mathbf{f}}_{\mathcal{X}}
\end{equation}
with
%
\begin{equation}
\label{Eq:Lestimation}
\mathbf{L}_{\mathcal{X}} = \mathbf{I} + \sum_{c=1}^{C} N_{\mathcal{X}}^{(c)}
\bar{\mathbf{T}}^{(c)'}  \bar{\mathbf{T}}^{(c)} ,
\end{equation}
where $\bar{\mathbf{f}}_{\mathcal{X}}^{(c)}$ and $\bar{\mathbf{T}}^{(c)}$
are the ``normalized'' variants of $\mathbf{f}_{\mathcal{X}}^{(c)}$ and $\mathbf{T}^{(c)}$, respectively:
\begin{eqnarray}
\label{Eq:NormF}
\bar{\mathbf{f}}_{\mathcal{X}}^{(c)}  & = & \EE^{(c)-\frac{1}{2}} \left(\mathbf{f}_{\mathcal{X}}^{(c)} -  N_{\mathcal{X}}^{(c)} \mm^{(c)} \right ) \\[-1mm]
\label{Eq:NormT}
\bar{\mathbf{T}}^{(c)}  & = & \EE^{(c)-\frac{1}{2}} \mathbf{T}^{(c)},
\end{eqnarray}
and $\EE^{(c)-\frac{1}{2}}$ is a symmetrical decomposition (such as Cholesky) 
of an inverse of the GMM UBM covariance matrix $\EE^{(c)}$.


\subsection{Factorization of i-vector Extractor}
\label{sec:DiX}

In this work, we propose to factorize each matrix $\tmatrix^{(c)}$ as:
%
\begin{eqnarray}
\bar{\tmatrix}^{(c)}  & = &  \sum^{Q}_{q=1} a_q^{(c)} \mathbf{U}_q,
\end{eqnarray}
where $Q$ is number of factors, $\mathbf{U}_q$ are the base matrices, $a_q^{(c)}$ are scalar weights for each component $\tmatrix^{(c)}$. Note that bases $\mathbf{U}_q$ are shared across all components $c$.  The number of parameters in this new model representation is $QC+QFD$, while the number parameters in the original i-vector
extractor was $CFD$.  Since there is no general requirement of linear independence for the individual matrices $\tmatrix^{(c)}$ in the original i-vector concept, the size of $Q$ would have to be equal to $C$ in order for the factorized model to fully describe the original subspace $\tmatrix$.  However, our assumption is that there, in fact, is some level of linear dependency and therefore, $Q$ can be chosen significantly smaller than $C$, therefore reducing the original model parameter space.


\subsection{Discriminatively Trained Factorized i-vector Extractor}
\label{ditrafiX}

\begin{figure}[tb]
  \centering
  \scalebox{1.0}{
  \includegraphics[width=1.0\linewidth]{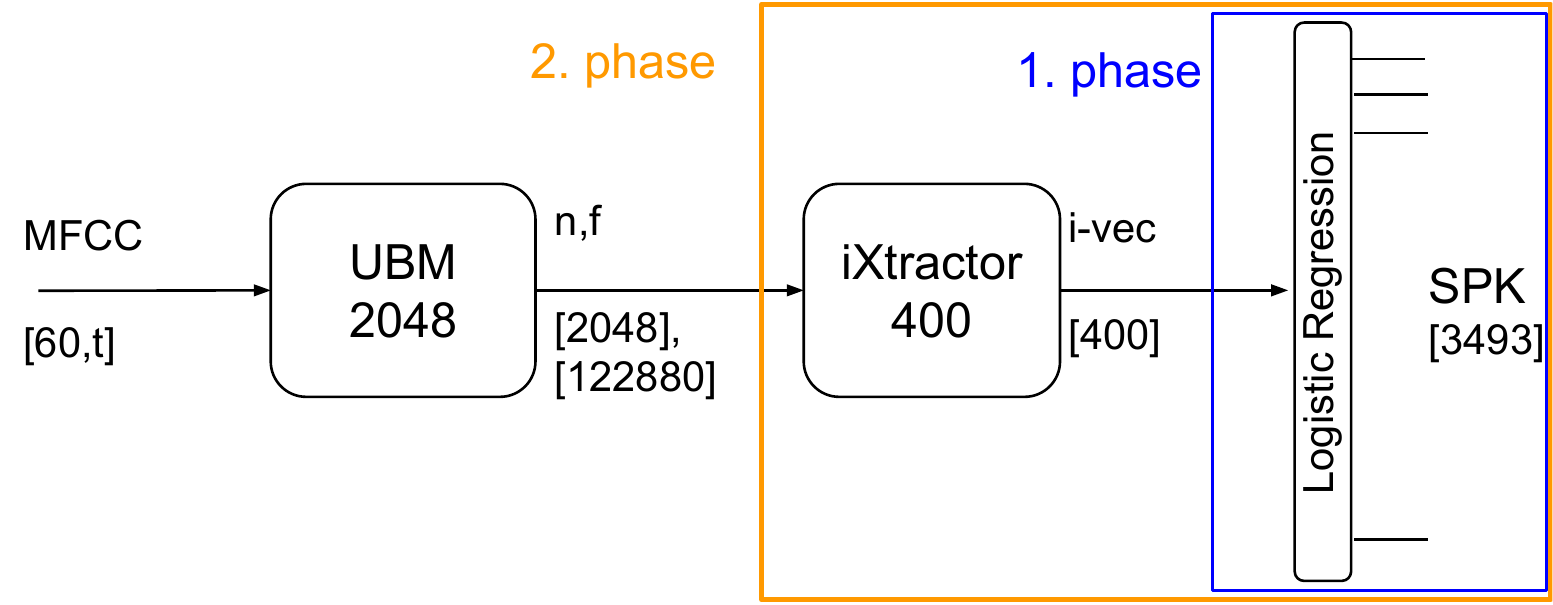}}
  \caption{Training pipeline of i-vector extractor parameters re-estimation. During the initial phase of training, only the logistic regression is trained. During the second phase, the parameters of the logistic regression and the i-vector extractor (iXtractor) are updated.}
  \label{fig:iXpipeline}
\end{figure}

In our previous work~\cite{novotny2018discriminatively}, discriminative training of $\mathbf{T}$ was based on using a multi-class logistic regression with parameters $\mathbf{W}$ as a classifier (classifying $K$ speakers), both being optimized based on the categorical cross entropy as an objective function (also depicted in Fig.~\ref{fig:iXpipeline}):
\begin{eqnarray}
\label{eq:prim_op_obj}
E(\mathbf{W}, \mathbf{T}) = -\sum^{N}_{n=1} \sum^{K}_{k=1} s_{nk} \log p_{\mathbf{W}}(C_{k} \mid \bphi_{\mathcal{X}_n}),
\end{eqnarray}
where, $s_{nk}$ is $k$-th 
element of the target variable in 1-of-K coding, $K$ is the number of speakers 
(classes), $N$ is the number of training samples, and $p_{\mathbf{W}}(C_{k} \mid \bphi_{\mathcal{X}_n})$ is a posterior probability (parametrized by logistic regression $\mathbf{W}$) of speaker $C_k$ given $n$-th utterance.  For the purpose of this work, 
let us treat
the i-vector $\bphi_{\mathcal{X}_n}$ as a function of $\tmatrix$.

Generatively trained i-vector extractor was used as an initialization.
In this work, we continue using this framework with some adjustments.  Let us generalize the optimization objective by adding an $L_2$ regularizer:
\begin{eqnarray} 
\label{eq:sec_op_obj}
E_{\mathrm{reg}}(\mathbf{W}, \mathbf{T}) = E(\mathbf{W}, \mathbf{T}) + \lambda||\tmatrix,\tmatrix_{\mathrm{orig}}||,
\end{eqnarray}
where $||\tmatrix,\tmatrix_{\mathrm{orig}}||$ is a Euclidian distance between our factorized matrix $\tmatrix$, and the original generatively trained matrix $\tmatrix_{\mathrm{orig}}$. 

We used two training schemes which differ in initialization and in the $\lambda$ regularizing factor.
In scheme-1 initialization, we select $Q$ eigen-vectors (based on $Q$ largest eigen-values) of covariance matrix of the vectorized $\mathbf{T}^{(c)}$'s (C vectors of  $FD$-dimensionality). Parameters $a_q^{(c)}$ are computed as a solution of system of $Q$ equations $\bar{\mathbf{T}}^{(c)}=\sum^{Q}_{q=1} a_q^{(c)} \mathbf{U}_q$. 
For this scheme, we globally set $\lambda = 0$.
In phase-1 of this scheme, only classifier $\mathbf{W}$ is trained in several epochs, until convergence on a cross-validation set is reached.  Then, in phase-2, both the classifier $\mathbf{W}$ and the extractor (represented by $\mathbf{U}_q$ and $a_q^{(c)}$) are retrained until convergence on a cross-validation set is reached.

In scheme-2, we started with random initialization, and for the first epoch (phase-0), $\lambda$ was set to a large number ($10^5$ in our case).  After that, $\lambda$ was set to zero for the rest of the training, and phase-1 and phase-2 coppied those in scheme-1.

We experimented with different $\lambda$-progression schemes (exponential decreasing, lower stable $\lambda$ during whole training, etc.), however, we discovered that one epoch was enough to reach the minimal distance to the $\tmatrix_{\mathrm{orig}}$. More epochs or learning rate decreasing did not bring any significant improvement neither in $||\tmatrix,\tmatrix_{\mathrm{orig}}||$ nor in final EERs.  

In general, we used stochastic gradient descent algorithm for parameter optimization.


\section{System Setup}


\subsection{Datasets}

We used the PRISM~\cite{ferrer:sre11} training dataset definition without added 
noise or reverberation to train UBM and i-vector extractor. 
The  set comprises Fisher 1 and 2, Switchboard phase 2 and 3 and Switchboard cellphone phases 
1 and 2, along with a set of Mixer speakers. This includes the 66 held out 
speakers from SRE10 (see Section III-B5 of~\cite{ferrer:sre11}), and 965, 980, 485 and 310 speakers 
from SRE08, SRE06, SRE05 and SRE04, respectively. A total of 13,916 speakers 
are available in Fisher data and 1,991 in Switchboard data.

Two variants of gender-independent PLDA models were trained: one on the clean training data, the second 
included also artificially added different mixes of noises and reverberation. 
Artificially added noise and reverb segments totaled approximately $24000$ segments or $30\%$ of total number of clean segments for PLDA training, see details in Sec.~\ref{lab:pldaaugset}.

We evaluated our systems on the \emph{female} portions
of NIST SRE 2010~\cite{NIST_SRE:WWW} ({\it tel-tel}, {\it int-int} and {\it int-mic}) and PRISM ({\it prism,noi}, {\it prism,rev} and {\it prism,chn}, see section III.B of~\cite{ferrer:sre11}), where {\it tel-tel} and {\it prism,chn} represent telephone speech, {\it int-int} and {\it int-mic} interview speech and {\it prism,noi} with {\it prism,rev} represent artificially corrupted speech with noise and reverberation.

Additionally, we used the {\it Core-Core} condition from the SITW challenge---\emph{sitw-core-core}. SITW~\cite{SITW_evaluation_plan} dataset is a large collection of real-world data exhibiting speech from individuals across a wide array of challenging acoustic and environmental conditions.

We also test on NIST SRE 2016~\cite{NIST:SRE2016}, but we split the trial set by language into Tagalog ({\it sre16-tgl-f})  and Cantonese ({\it sre16-yue-f}).  We use only female trials (both single- and multi-session). We did not use SRE'16 unlabeled development set in any way.

We randomly selected 500 utterances from 500 different speakers as a cross-validation set from the PRISM training dataset.

\subsection{PLDA and i-vector Extractor Augmentation Sets}
\label{lab:pldaaugset}
To extend the training set, we created new artificially corrupted training sets from the PRISM training set. In addition to using noise and reverberation, data were also augmented with randomly generated cuts. In our experiments, we used 30\% of original training data to generate cuts with durations between 3 to 5 seconds.
The composition of the augmentation set is described in details in~\cite{xvec:ondran}.

\begin{table*}[!htb]
\centering
\caption{\label{tab:mainresults} 
Results in terms of EER $[\%]$ for different i-vector extractors: \ExpBas{} - generative baseline without augmented data, \ExpSchOneInit{} and \ExpSchTwoInit{} are mere initialized factorized models while \ExpSchOne{} and \ExpSchTwo{} are their re-trained variants. \ExpFull{} stands for a full representation of the original i-vector extractor that has been discriminatively re-trained.  The table is also verticaly divided into two blocks which correspond to the training set of PLDA, where we used either only clean data or multi-condition style of training (with noised and reverberated data added to the training of PLDA).}

\vspace{2mm}
\scalebox{0.8}{
\begin{tabular}{ l  r r r r r r  c  r r r r r r}
\toprule
\addlinespace[0.05cm]

\multicolumn{1}{c}{}  & 
\multicolumn{6}{c}{PLDA clean} & & \multicolumn{6}{c}{PLDA extension data} \\

\cmidrule(rl){2-7} \cmidrule(rl){9-14} 

Condition & 
  \multicolumn{1}{c}{\ExpBas{}} & \multicolumn{1}{c}{\ExpSchOneInit{}} & \multicolumn{1}{c}{\ExpSchOne{}} & \multicolumn{1}{c}{\ExpSchTwoInit{}} & \multicolumn{1}{c}{\ExpSchTwo{}} & \multicolumn{1}{c}{\ExpFull{}} & &
  \multicolumn{1}{c}{\ExpBas{}}    & \multicolumn{1}{c}{\ExpSchOneInit{}} & \multicolumn{1}{c}{\ExpSchOne{}} & \multicolumn{1}{c}{\ExpSchTwoInit{}} & \multicolumn{1}{c}{\ExpSchTwo{}} & \multicolumn{1}{c}{\ExpFull{}} \\

\midrule
tel-tel & 2.23 & 8.39 & 3.9 & 2.47 & 2.2 & 1.97 & & 3.36 & 9.72 & 4.91 & 3.52 & 3.3 & 3.25 \\
sre16-yue-f & 10.9 & 17.39 & 12.79 & 11.29 & 10.96 & 10.97 & & 11.32 & 17.18 & 12.18 & 11.42 & 11.11 & 10.87 \\
\midrule
int-int & 4.72 & 9.56 & 5.57 & 4.74 & 4.51 & 4.37 & & 4.83 & 10.18 & 5.94 & 4.96 & 4.67 & 4.56 \\ 
int-mic & 2.15 & 5.27 & 2.69 & 2.23 & 2.18 & 2.11 & & 2.02 & 5.67 & 2.65 & 2.28 & 2.1 & 1.91 \\
prism,chn & 1.13 & 5.63 & 2.25 & 0.92 & 0.83 & 0.88 & & 1.14 & 5.95 & 1.98 & 1.11 & 1.12 & 1.14 \\
sitw-core-core & 10.51 & 17.97 & 12.35 & 10.92 & 10.4 & 10.29 & & 10.57 & 17.54 & 12.33 & 10.84 & 10.47 & 10.21 \\
\midrule
prism,noi & 4.34 & 11.74 & 6.15 & 4.6 & 4.29 & 3.97 & & 3.66 & 10.73 & 5.27 & 4.04 & 3.73 & 3.44 \\
prism,rev & 2.81 & 8.59 & 3.67 & 2.84 & 2.49 & 2.54 & & 2.45 & 7.25 & 3.17 & 2.49 & 2.3 & 2.34 \\

\bottomrule
\end{tabular}}
\end{table*}


\section{Experiments and Discussion}
\label{sec:experiments}

One of the issues we had to solve to even begin experimenting with the factorized model was its proper initialization. We present two different strategies for initialization and then we will experiment with subsequent discriminative retraining of such models.  We also provide comparisons with the generative baseline and with discriminative retraining of its full representation.
In our experiments with factorization, we set the number of bases $Q$ to 250. This means that the matrix $\tmatrix$ is represented by 7.5 times less parameters compared to the original model $\tmatrix_{\mathrm{orig}}$, and when compared to the i-vector extractor block from from~\cite{rohdin:icassp:2018}) in Fig.~\ref{fig:end2end}, the number of parameters is almost half.  In all of our experiments, we set the i-vector subspace dimensionality to 400.

For clarity, we denote different ways of obtaining the i-vector extractor by capital letter \ExpBas{}, \ExpSchOne{}, \ExpSchTwo{} and \ExpFull{}:
\begin{description} 
\item [\ExpBas{}]  We trained a baseline i-vector extractor in the traditional generative way, using the original PRISM training corpus without any augmentations. 
\item [\ExpSchOneInit{}]  We initialized the bases $\mathbf{U}_b$ for factorized model by eigen-vectors. 

\item [\ExpSchOne{}]  We initialized the bases $\mathbf{U}_b$ for factorized model by eigen-vectors as in \ExpSchOneInit{} and then we continued training with the loss function from (\ref{eq:prim_op_obj}) and the two stage training described in Sec.~\ref{ditrafiX}. 
\item[\ExpSchTwoInit{}]  We initialized the bases $\mathbf{U}_b$ randomly and then we ran a single epoch of training with the loss function in (\ref{eq:sec_op_obj}). 

\item[\ExpSchTwo{}]  We initialized the bases $\mathbf{U}_b$ randomly and then we ran a single epoch of training with the loss function in (\ref{eq:sec_op_obj}) as in \ExpSchTwoInit{}, and then we continued training with the loss function from (\ref{eq:prim_op_obj}) and the two stage training described in Sec.~\ref{ditrafiX}.

\item[\ExpFull]  We driscriminatively re-trained a full representation of the baseline generative i-vector extractor~\cite{novotny2018discriminatively}.
\end{description}
To avoid over-fitting of the classifier during discriminative training, it was necessary to filter the training data. We selected speakers with at least 5 utterances in the original data.  This step limits the training data to 3493 speakers with 59112 utterances (177336 utterances including augmentation). 

For all experiments, we kept the same PLDA configuration. The i-vectors are pre-processed with mean normalization, LDA (i-vectors are transformed into 200-dimensions) and finally, they are length normalized.

Our results in terms of EER are presented in Tab.~\ref{tab:mainresults} which is divided into two vertical blocks to provide a comparison between PLDA trained on the clean data and multi-condition PLDA training, where we train the PLDA also on augmented copies of its training data. We are now interested in the general robustness of our methods and therefore we will focus on overall performance across all conditions rather than looking closely into individual cases. 

The table is also divided into three horizontal blocks based on the type of the condition: into telephone channel ({\it tel-tel}, {\it sre16-yue-f}), microphone ({\it int-int}, {\it int-mic}, {\it prism,ch}, {\it sitw-core-core}) and artificially created conditions ({\it prism,noi}, {\it prism,rev}). We did not use any type of adaptation, score normalization or any other technique used for results improvement in conditions from SRE16 and others. For system \ExpSchOne{} and \ExpSchTwo{}, we also present results for initialization, before $\mathbf{U}_b$ were retrained (in \ExpSchTwo{} after first epoch with $\lambda||\tmatrix,\tmatrix_{\mathrm{orig}}||$ penalty).

When we compare baseline systems (columns \ExpBas{} in the table) with the results obtained with initialized models for discriminative training (columns \ExpSchOneInit{} and \ExpSchTwoInit{}), we can see that \ExpSchOneInit{} is always significantly worse than the baseline. Initialization \ExpSchTwoInit{} has much better results as they are only slightly degraded compared to the baseline indicating that we were able to represent the original i-vector model well.

We can see, that starting from \ExpSchOneInit{}, we reach significant improvements with discriminative parameters re-estimation.  Unfortunately, these results indicate, that the model got stuck in the local minimum and it was not able to improve to the level of the baseline. 

Initialization variant \ExpSchTwoInit{} proved to be a significantly better starting point. After discriminative parameter re-estimation, the model \ExpSchTwo{} was able to obtain slight improvement across all conditions w.r.t. \ExpSchTwoInit{}. Model \ExpSchTwo{} has also achieved a slight improvement over the baseline \ExpBas{} or almost reached its performance.

Observing results in columns \ExpFull{}, we can compare with discriminative retraining of the full i-vector representation~ \cite{novotny2018discriminatively}. With model \ExpFull{} we achieve the best overall performance (slightly better than \ExpSchTwo{}), but the architecture with factorization offers approximately 4 times faster training with 7.5 times less parameters which will allow us to further extend the model and include also the GMM-UBM representation.

\section{Conclusion}
In this work, we have presented a way of refining a discriminative training of i-vector extractor from our previous work.  We were able to slightly outperform the generative baseline.
Our approach conveniently fits to the current efforts of building a fully end-to-end discriminative systems, and provides a way for a robust initialization of such a large and important part of the system. Needless to say, we have not created a new state-of-the-art system, however, we have prepared a solid platform for our further research. In our ongoing research, we will focus on the final close-form solution of the generative objective and direct estimation of $\mathbf{U}_b$, which will be helpful for simpler initialization.  We also plan to analyze the effect of number of factors $Q$.

\bibliographystyle{IEEEtran}

\bibliography{template}

\begin{thebibliography}{10}
\providecommand{\url}[1]{#1}
\csname url@samestyle\endcsname
\providecommand{\newblock}{\relax}
\providecommand{\bibinfo}[2]{#2}
\providecommand{\BIBentrySTDinterwordspacing}{\spaceskip=0pt\relax}
\providecommand{\BIBentryALTinterwordstretchfactor}{4}
\providecommand{\BIBentryALTinterwordspacing}{\spaceskip=\fontdimen2\font plus
\BIBentryALTinterwordstretchfactor\fontdimen3\font minus
  \fontdimen4\font\relax}
\providecommand{\BIBforeignlanguage}[2]{{%
\expandafter\ifx\csname l@#1\endcsname\relax
\typeout{** WARNING: IEEEtran.bst: No hyphenation pattern has been}%
\typeout{** loaded for the language `#1'. Using the pattern for}%
\typeout{** the default language instead.}%
\else
\language=\csname l@#1\endcsname
\fi
#2}}
\providecommand{\BIBdecl}{\relax}
\BIBdecl

\bibitem{lozano_odyssey_2016}
\BIBentryALTinterwordspacing
A.~Lozano-Diez, A.~Silnova, P.~Mat{\v{e}}jka, O.~Glembek, O.~Plchot,
  J.~Pe{\v{s}}{\'{a}}n, L.~Burget, and J.~Gonzalez-Rodriguez,
  ``\BIBforeignlanguage{english}{{A}nalysis and {O}ptimization of {B}ottleneck
  {F}eatures for {S}peaker {R}ecognition},'' in
  \emph{\BIBforeignlanguage{english}{Proceedings of Odyssey 2016}}, vol. 2016,
  no.~06.\hskip 1em plus 0.5em minus 0.4em\relax International Speech
  Communication Association, 2016, pp. 352--357. [Online]. Available:
  \url{http://www.fit.vutbr.cz/research/view_pub.php.cz.iso-8859-2?id=11219}
\BIBentrySTDinterwordspacing

\bibitem{Lei_icassp_2014}
Y.~Lei, N.~Scheffer, L.~Ferrer, and M.~McLaren, ``A novel scheme for speaker
  recognition using a phonetically-aware deep neural network,'' in \emph{2014
  IEEE International Conference on Acoustics, Speech and Signal Processing
  (ICASSP)}, May 2014, pp. 1695--1699.

\bibitem{Novoselov_interspeech_2015}
S.~Novoselov, T.~Pekhovsky, O.~Kudashev, V.~S. Mendelev, and A.~Prudnikov,
  ``{N}on-linear {PLDA} for i-vector speaker verification,'' in \emph{2017 IEEE
  International Conference on Acoustics, Speech and Signal Processing
  (ICASSP)}, Sept 2015, pp. 214--218.

\bibitem{Bhattacharya_SLT16}
G.~Bhattacharya, J.~Alam, P.~Kenny, and V.~Gupta, ``Modelling speaker and
  channel variability using deep neural networks for robust speaker
  verification,'' in \emph{2016 {IEEE} Spoken Language Technology Workshop,
  {SLT} 2016, San Diego, CA, USA, December 13-16}, 2016.

\bibitem{Ghahabi_icassp_2014}
O.~Ghahabi and J.~Hernando, ``Deep belief networks for i-vector based speaker
  recognition,'' in \emph{2014 IEEE International Conference on Acoustics,
  Speech and Signal Processing (ICASSP)}, May 2014, pp. 1700--1704.

\bibitem{Variani_icassp_2014}
E.~Variani, X.~Lei, E.~McDermott, I.~L. Moreno, and J.~Gonzalez-Dominguez,
  ``Deep neural networks for small footprint text-dependent speaker
  verification,'' in \emph{2014 IEEE International Conference on Acoustics,
  Speech and Signal Processing (ICASSP)}, May 2014, pp. 4052--4056.

\bibitem{heighold_icassp_2016}
G.~Heigold, I.~Moreno, S.~Bengio, and N.~Shazeer, ``End-to-end text-dependent
  speaker verification,'' in \emph{2016 IEEE International Conference on
  Acoustics, Speech and Signal Processing (ICASSP)}, March 2016, pp.
  5115--5119.

\bibitem{zhang_slt_2016}
S.~X. Zhang, Z.~Chen, Y.~Zhao, J.~Li, and Y.~Gong, ``{E}nd-to-{E}nd attention
  based text-dependent speaker verification,'' in \emph{2016 IEEE Spoken
  Language Technology Workshop (SLT)}, Dec 2016, pp. 171--178.

\bibitem{snyder_slt_2016}
D.~Snyder, P.~Ghahremani, D.~Povey, D.~Garcia-Romero, Y.~Carmiel, and
  S.~Khudanpur, ``{D}eep neural network-based speaker embeddings for end-to-end
  speaker verification,'' in \emph{2016 IEEE Spoken Language Technology
  Workshop (SLT)}, Dec 2016, pp. 165--170.

\bibitem{Bhattacharaya_interspeech_2017}
G.~Bhattacharya, J.~Alam, and P.~Kenny, ``{D}eep {S}peaker {E}mbeddings for
  {S}hort-{D}uration {S}peaker {V}erification,'' in \emph{Interspeech 2017}, 08
  2017, pp. 1517--1521.

\bibitem{snyder_interspeech_2017}
D.~Snyder, D.~Garcia-Romero, D.~Povey, and S.~Khudanpur, ``{D}eep {N}eural
  {N}etwork {E}mbeddings for {T}ext-{I}ndependent {S}peaker {V}erification,''
  in \emph{Interspeech 2017}, Aug 2017.

\bibitem{xvec:Snyder2018}
D.~Snyder, D.~Garcia-Romero, G.~Sell, D.~Povey, and S.~Khudanpur,
  ``{X}-vectors: {R}obust {DNN} {E}mbeddings for {S}peaker {R}ecognition,'' in
  \emph{Proceedings of ICASSP}, 2018.

\bibitem{novotny2018discriminatively}
O.~Novotny, O.~Plchot, O.~Glembek, L.~Burget, and P.~Matejka,
  ``Discriminatively re-trained i-vector extractor for speaker recognition,''
  \emph{accepted to ICASSP 2019}, 2019.

\bibitem{Axelrod2003LargeVC}
S.~Axelrod, V.~Goel, B.~Kingsbury, K.~Visweswariah, and R.~A. Gopinath, ``Large
  vocabulary conversational speech recognition with a subspace constraint on
  inverse covariance matrices,'' in \emph{INTERSPEECH}, 2003.

\bibitem{PoveySPAM}
\BIBentryALTinterwordspacing
D.~Povey, ``{SPAM} and full covariance for speech recognition,'' in
  \emph{{INTERSPEECH} 2006 - ICSLP, Ninth International Conference on Spoken
  Language Processing, Pittsburgh, PA, USA, September 17-21, 2006}, 2006.
  [Online]. Available:
  \url{http://www.isca-speech.org/archive/interspeech\_2006/i06\_2047.html}
\BIBentrySTDinterwordspacing

\bibitem{sandroFACTOR}
S.~Cumani and P.~Laface, ``Factorized sub-space estimation for fast and memory
  effective i-vector extraction,'' \emph{Audio, Speech, and Language
  Processing, IEEE/ACM Transactions on}, vol.~22, pp. 248--259, 2014.

\bibitem{dix:glembek}
\BIBentryALTinterwordspacing
O.~Glembek, L.~Burget, N.~Br{\"{u}}mmer, O.~Plchot, and P.~Mat{\v{e}}jka,
  ``\BIBforeignlanguage{english}{{D}iscriminatively {T}rained i-vector
  {E}xtractor for {S}peaker {V}erification},'' in
  \emph{\BIBforeignlanguage{english}{Proceedings of Interspeech 2011}},
  no.~8.\hskip 1em plus 0.5em minus 0.4em\relax International Speech
  Communication Association, 2011, pp. 137--140. [Online]. Available:
  \url{http://www.fit.vutbr.cz/research/view_pub.php.cs?id=9752}
\BIBentrySTDinterwordspacing

\bibitem{xvec:ondran}
\BIBentryALTinterwordspacing
O.~Novotn{\'{y}}, O.~Plchot, P.~Mat{\v{e}}jka, L.~Mo{\v{s}}ner, and O.~Glembek,
  ``\BIBforeignlanguage{english}{{O}n the use of {X}-vectors for {R}obust
  {S}peaker {R}ecognition},'' in \emph{\BIBforeignlanguage{english}{Proceedings
  of Odyssey 2018}}, no.~6.\hskip 1em plus 0.5em minus 0.4em\relax
  International Speech Communication Association, 2018, pp. 168--175. [Online].
  Available: \url{http://www.fit.vutbr.cz/research/view_pub.php.cs?id=11787}
\BIBentrySTDinterwordspacing

\bibitem{rohdin:icassp:2018}
J.~Rohdin, A.~Silnova, M.~Diez, O.~Plchot, P.~Mat\v{e}jka, and L.~Burget,
  ``{E}nd-to-end {DNN} based speaker recognition inspired by i-vector and
  {PLDA},'' in \emph{Proceedings of ICASSP}.\hskip 1em plus 0.5em minus
  0.4em\relax IEEE Signal Processing Society, 2018.

\bibitem{DehakN_TASLP:2010}
N.~Dehak, P.~Kenny, R.~Dehak, P.~Dumouchel, and P.~Ouellet, ``Front-{E}nd
  {F}actor {A}nalysis {F}or {S}peaker {V}erification,'' \emph{IEEE Transactions
  on Audio, Speech, and Language Processing}, vol.~19, no.~4, pp. 788--798, May
  2011.

\bibitem{ferrer:sre11}
L.~Ferrer, H.~Bratt, L.~Burget, H.~Cernocky, O.~Glembek, M.~Graciarena,
  A.~Lawson, Y.~Lei, P.~Matejka, O.~Plchot, and N.~Scheffer, ``{P}romoting
  robustness for speaker modeling in the community: the {PRISM} evaluation
  set,'' in \emph{Proceedings of {SRE11} analysis workshop}, Atlanta, Dec.
  2011.

\bibitem{NIST_SRE:WWW}
``{N}ational {I}nstitute of {S}tandards and {T}echnology,''
  http://www.nist.gov/speech/tests/spk/index.htm.

\bibitem{SITW_evaluation_plan}
\BIBentryALTinterwordspacing
M.~McLaren, L.~Ferrer, D.~Castan, and A.~Lawson, ``{T}he {S}peakers in the
  {W}ild ({SITW}) {S}peaker {R}ecognition {D}atabase,'' in \emph{Interspeech
  2016}, 2016, pp. 818--822. [Online]. Available:
  \url{http://dx.doi.org/10.21437/Interspeech.2016-1129}
\BIBentrySTDinterwordspacing

\bibitem{NIST:SRE2016}
``The {NIST} year 2016 {S}peaker {R}ecognition {E}valuation {P}lan,'' \url{
  https://www.nist.gov/sites/default/files/documents/2016/10/\\07/sre16\_eval\_plan\_v1.3.pdf},
  2016.

\end{thebibliography}

\end{document}